\documentclass[a4paper,onecolumn, 11pt]{quantumarticle}
\pdfoutput=1
\usepackage[utf8]{inputenc}
\usepackage[english]{babel}
\usepackage[T1]{fontenc}
\usepackage{amsmath}
\usepackage{hyperref}

\usepackage{tikz}
\usepackage{lipsum}

\newcommand{\bra}[1]{\ensuremath{\langle{#1}|}}
\newcommand{\ket}[1]{\ensuremath{|{#1}\rangle}}

\DeclareMathOperator{\Tr}{Tr}

\begin{document}

\title{Optimal quantum sensing of the nonlinear bosonic interactions using Fock states}

\author{Payman Mahmoudi}
\affiliation{J. Heyrovsk{\'y} Institute of Physical Chemistry of the ASCR, v.v.i., Dolej{\v s}kova 2155/3, 182 23 Prague 8, Czech Republic.}
\affiliation{Department of Optics, Palacky University, 17. listopadu 12, 771 46 Olomouc, Czech Republic.}
\email{payman.mahmoudi@jh-inst.cas.cz}
\orcid{0000-0003-2750-4738}

\author{Atirach Ritboon}
\affiliation{Department of Optics, Palacky University, 17. listopadu 12, 771 46 Olomouc, Czech Republic.}
\affiliation{Research Unit in Energy Innovations and Modern Physics (EIMP), Thammasat University, Khlong Nueng, Khlong Luang, Pathum Thani, 12120, Thailand.}
\email{ritboon@tu.ac.th}
\orcid{0000-0003-2519-748X}

\author{Radim Filip}
\email{filip@optics.upol.cz}
\orcid{0000-0003-4114-6068}
\affiliation{Department of Optics, Palacky University, 17. listopadu 12, 771 46 Olomouc, Czech Republic.}
\maketitle

\begin{abstract}
Nonlinear processes with individual quanta beyond bilinear interactions are essential for quantum technology with bosonic systems. Diverse coherent splitting and merging of quanta in them already manifest in the estimation of their nonlinear coupling from observed statistics. We derive non-trivial, but optimal strategies for sensing the basic and experimentally available trilinear interactions using non-classical particle-like Fock states as a probe and feasible measurement strategies. Remarkably, the optimal probing of nonlinear coupling reaches estimation errors scaled down with $N^{-1/3}$ for overall $N$ of quanta in specific but available high-quality Fock states in all interacting modes. It can reveal unexplored aspects of nonlinear dynamics relevant to using such nonlinear processes in bosonic experiments with trapped ions and superconducting circuits and opens further developments of quantum technology with them. 
\end{abstract}

\section{Introduction}
The nonlinear interactions have been at the center of interest in the fields of quantum optics, quantum processing and even those beyond \cite{ Bloembergen1996, Louisell1961, Perina1991, Braunstein2005, Dorfman2016, Nation2010} since its emergence in the discovery of the first second-harmonic generation by Franken {\it et al}. \cite{Franken1961}, shortly after the first introduction of practical lasers \cite{Maiman1960}. Its studies induce both fundamental understanding \cite{Nimmrichter2017, Agarwal1988, Mista2001} explaining interesting, sometimes unexpected, features and behaviors of the involved systems\cite{Laha2022, Xing2022, Levy2012} and technological applications such as telecommunications \cite{Schneider2004,Quemard2001}, laser technologies \cite{Brabec2000, Jhon2020}, spectroscopy \cite{Fischer2005, Dorfman2016}, two-photon imaging \cite{Abouraddy2001, Pittman1995}, and remarkably those in which the non-classical effects are exploited to the core \cite{Dorfman2016, Mitchison2019}. Quantum technologies using quantum nonlinear interactions are a stimulating research area and become ubiquitous. For example, optical non-linearity at the level of individual photons in strong photon-photon interactions allows us to perform quantum-by-quantum control of light fields and single-photon switches \cite{Chang2014}. Spontaneous parametric down-conversion generated by a nonlinear interaction in a nonlinear crystal is another example of an entanglement source \cite{Couteau2018, Zhang2021, Franke-Arnold2002}. The down-converted photon pairs are utilized in various protocols of quantum processing, especially in quantum cryptography \cite{Couteau2018,Sergienko1999, Naik2000}. Moreover, it has been demonstrated that nonlinearity is essential for implementing universal quantum gates for continuous-variable quantum computation (CVQC)\cite{Lloyd1999,Yanagimoto2023}, the key for reprogrammable bosonic quantum logic gates \cite{Krastanov2021} and photon-number-resolving quantum nondemolition \cite{Yanagimoto2023}.

Trilinear coupling between three different bosonic fields, in particular, often plays an important role in quantum optics for a long time as it gives physical descriptions for many interesting optical processes such as frequency conversion, Raman and Brillouin scattering, parametric amplification, parametric oscillation in optical cavities \cite{Walls1970, Drobny1992, Mishkin1969, Walls1972, McNeil1983}. It also brings many applications into the field such as quantum demolition measurement \cite{Gagen1991}, generation of super-radiant states \cite{Liu1991}, SU(1,1) and SU(2) squeezing \cite{Jex1992}.  This type of interactions is experimentally available in ion-trapping platforms, using the anharmonicity of the Coulomb potential \cite{Ding2018, Maslennikov2019} and in superconducting circuit platforms \cite{Chang2020, Vrajitoarea2020}. Each bosonic field modes are realized by the mechanical motion of the three atoms in trapped ion platforms. Recently, this coupling has been utilized to effectively demonstrate quantum simulation \cite{Ding2018} and quantum refrigerator \cite{Maslennikov2019, Nimmrichter2017}, which paves the way to study the thermodynamics of single quantum systems and allows us to perform a non-Gaussian gate of CVQC. In addition, this nonlinearity is employed to perform phonon counting in trapped ions \cite{Ding2017}. The scientific interest in this type of interactions is not limited only to the optical and ion-trapping communities.  In the field of quantum gravitation, it is used to semiclassically explain the Hawking radiation of a black hole \cite{Nation2010, Alsing2015, Bradler2016}.

On the other hand, quantum metrology aims to explore optimal quantum strategies and protocols to estimate an unknown parameter $\theta$ with the smallest possible uncertainty $\Delta\theta$ through the change of the probability distribution of the probe \cite{Paris2009, Giovannetti2011, Toth2014, Pezze2018}. With the probabilistic nature of quantum measurements, the inference of the parameter $\theta$ cannot be accomplished by a single-short measurement but through the measurement of an ensemble in a scale of thousands or even millions of particles. Quantum-enhanced metrology can be achieved by improving the sensing procedures, such as enhancing the sensitivity using the nonclassicality of the probe \cite{Pezze2018, Jones2009, Wolf2019}, optimizing the interaction time \cite{Florez2018}, or using effective quantum measurement \cite{Wiseman1995, Higgins2007, Ritboon2022}. For example, a superposition between a ground state $\ket{0}$ and a number state $\ket{n}$ of atomic motion can be used to optimally estimate the changes in frequencies of mechanical oscillators \cite{McCormick2019}. A motional Fock state, on the other hand, has been reported to be an optimal state for sensing a phase-randomized displacement in trapped ions \cite{Wolf2019}. A Gaussian squeezed vacuum state has been demonstrated to be optimal for optical phase sensing using homodyne detection \cite{Nielsen2023}.

 To witness the nonlinear interaction for further advancements and exploit its full potential, its initially small coupling strength has to be sensitively measured to quantify the interaction rate. The smallest possible value of the coupling strength can be measured, so the more new interactions and applications to be revealed. However, as the coupling strength is not a quantum observable, its estimation has to be accomplished through quantum metrology by observing the statistics of an observable. One of the approaches for sensing nonlinear coupling strength is recently proposed in \cite{Ivanov2022}. This scheme theoretically shows that in a weak coupling regime, a linear coupling between a spin qubit and a motional mode nonlinearly coupled with the other mode can be reduced into a spin-dependent motional squeezing or a spin-dependent beam splitter operator. The estimated coupling strength of the nonlinear interaction thus can be determined through the sensing of these parameters by measuring the phonon distribution using adiabatic transition, proposed theoretically in \cite{Kirkova2021}. 

Our research question, however, is intrinsically different from that of \cite{Ivanov2022}, as we aim not only to design the sensing protocol but also to investigate the optimal quantum state of the motional modes, used as the probe, for sensing the coupling strength of the nonlinear interactions. The nonlinear interaction between the motional modes is operated as the sensing process before using a qubit as an ancillary to measure the change in the phonon distribution due to the interaction. The technique of using qubits for measuring the population of phonons, through Jaynes-Cumming interaction, is rather well established in the field of trapped ions \cite{Leibfried2003}. We find that, for small values of the coupling strength, once all interacting bosonic modes are prepared in specific Fock states, where all interacting modes have their energy quanta with the same ratio as the number of annihilation and creation operators associated with the interacting modes in the interaction Hamiltonian. This will give the optimal sensing with the estimation error proportional to $N^{-1/3}$, where $N$ is the total bosonic excitation number.

\section{Overview}
In this paper, we consider two different types of trilinear interactions, whose Hamiltonian are given explicitly in the following section. The first interaction couples three bosonic modes in such a way that simultaneous absorption of  excitation from two interacting modes, modes $b$ and $c$, excites the other mode, mode $a$. Reversely, an absorption of mode $a$, in return, provides a quanta to modes $b$ and $c$ each. For brevity, we name this interaction as interaction I. The second interaction, on the other hand, describes the coupling between two distinct bosonic modes. An absorption of two excitations in one mode (mode $b'$) would give a quanta of energy to the other mode (mode $a'$), and we name this later interaction as interaction II.

\begin{figure}
    \centering
    \includegraphics[width=\textwidth]{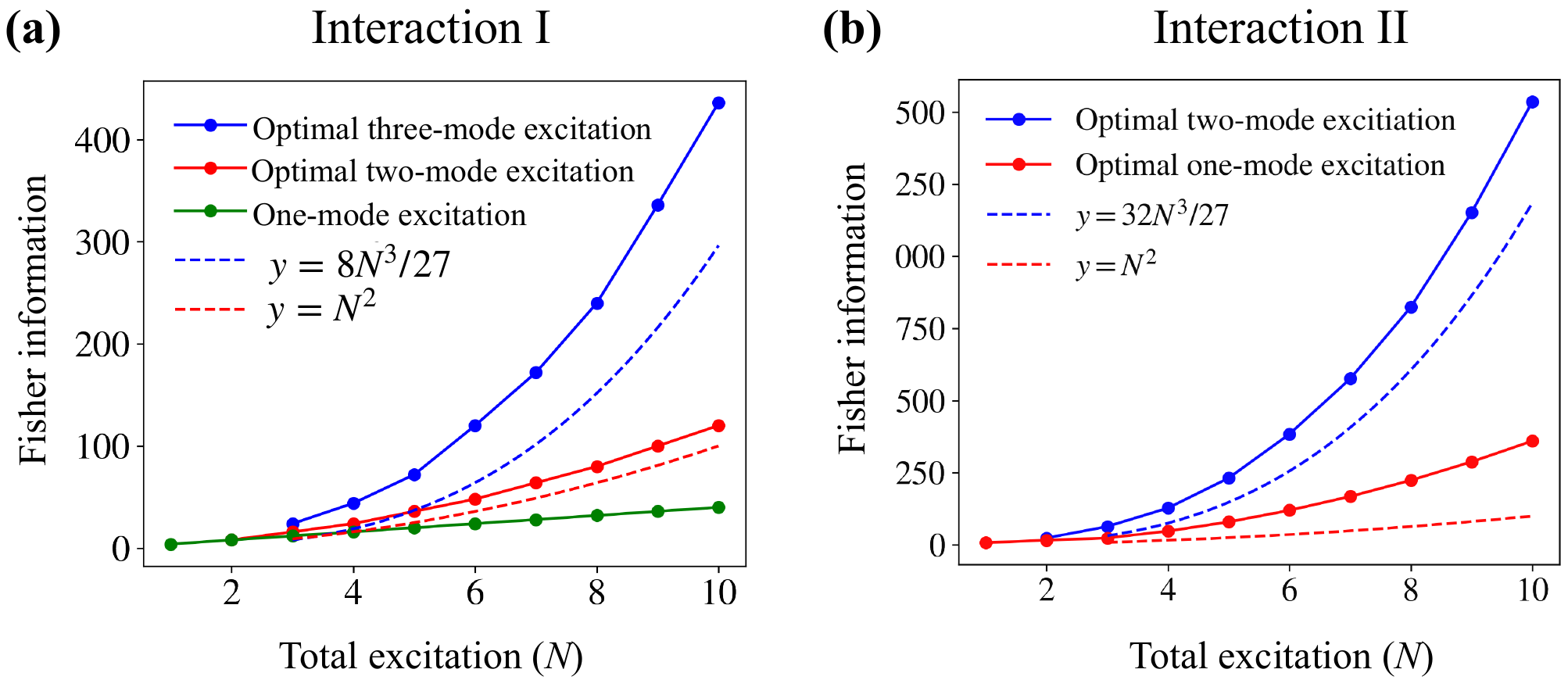}
    \caption{The Fisher information for sensing small values of coupling constants, $\xi, \chi\ll1$, when the motional modes are prepared in optimal configurations of Fock states, increases with the total phonon number $N$ differently depending on the approaches of the state preparation. {(\bf a)} For the non-degenerate interaction I, Eq.\ref{eq:1}, the graph of the figure illustrates cubic, quadratic and linear relations between the Fisher information and $N$ for sensing the coupling constant $\xi$ using optimal configurations of three-mode, two-mode and single-mode excitation approaches respectively. {\bf (b)} For degenerate interaction II, Eq. \ref{eq:2}, on the other hand, using optimal configurations of the probe gives cubic and quadratic relations between Fisher information and total phonon number $N$ for two- and single-mode excitation approaches. The dashed lines represent the first-order approximation of the optimal Fisher information using different excitation schemes.}
    \label{fig:6}
\end{figure}

The proposed sensing scheme for probing the coupling strengths $\chi$ and $\xi$ of interactions I and II respectively is given in detail in section \ref{sec:protocol}. In short, the scheme includes three different stages: probe preparation, sensing and measurement. For the probe preparation stage, we suggest that the interacting motional modes, regarded as a probe, must be optimally prepared in specific tensors of Fock states. Subsequently, the trilinear interactions are resonantly run for a short time, such that $\chi t<1$ (or $\xi t<1$), so that the probe's state is not severely disturbed to an extent that the coupling strength can no longer be implied uniquely. Finally, we measure the change in the populations of the probe by coupling one of the modes with a qubit and performing a projective measurement on the qubit.

The high sensitivity of the protocol for sensing these interactions originates from both the efficient ability to detect small deviations in the phonon population of the probe initially in Fock states through qubit coupling and the sensitivity of Fock states due to their unique nonclassical nature \cite{Wolf2019}. The probe in an ideal Fock state $\ket{n}$, by definition, does not have any population in any other state initially. Therefore, a small change in the population distribution thus directly implies the effect of the nonlinear interactions during the sensing process. Moreover, high-energy Fock states, with an excitation larger than 10 quanta, have been successfully prepared using both trapped ions \cite{Podhora2022} and superconducting microwave cavities \cite{Deng2022}. For sensing the coupling strength of interactions I and II, the best configuration of Fock states is achieved once all interacting motional modes share their excitation in the same ratio as the number of operators associated with the modes in the interaction Hamiltonian. The sensitivity, quantified by classical Fisher information, scales up with $N^3$ where $N$ is the total excitation of the motional modes as indicated by the blue solid-dotted lines in figure \ref{fig:6}.  With the Cramer-Rao bound \cite{Cramer1999,Rao1992}, the estimation error per experimental run of the couplings scales down by $N^{-3}$, as a result. The red and green lines of the figure represent the optimal sensitivities in the scenarios when one (red) or two (green) of the interacting motional modes are set in their ground state respectively. The dashed lines are displayed to compare them with the polynomial rise of sensitivities. This reflects that the nonlinear interactions induce the disturbance in the population distribution of a Fock state proportional to the product of the excitations in each interacting mode, shown in the appendix.  

\section{The nonlinear interactions}
Interaction I is a trilinear interaction of three different motional modes of the trapped atoms caused by the an-harmonic Coulomb repulsion between them. The Hamiltonian of the non-degenerate interaction can be expressed as \cite{Ding2018, Maslennikov2019}
\begin{align}\label{eq:1}
    \hat{H}_{\rm I}=\hbar\omega_a\hat{a}^\dagger\hat{a}+\hbar\omega_b\hat{b}^\dagger\hat{b}+\hbar\omega_c\hat{c}^\dagger\hat{c}+\hbar\xi\left(\hat{a}^\dagger\hat{b}\hat{c}+\hat{a}\hat{b}^\dagger\hat{c}^\dagger\right),
\end{align}
where where $\hat{a} (\hat{a}^\dagger)$, $\hat{b} (\hat{b}^\dagger)$, and $\hat{c} (\hat{c}^\dagger)$ are the annihilation (creation) operators for the motional modes $a$, $b$, and $c$ of the three ions whose oscillation frequencies are $\omega_a$, $\omega_b$ and $\omega_c$ respectively, and $\xi$ is denotes the interaction strength, which is the parameter to be sensed and quantified. In this work, we consider only the case when the resonance condition, of the form $\omega_a=\omega_b+\omega_c$, is met for non-degenerate quanta at frequencies $\omega_b$ and $\omega_c$. The Hamiltonian in this condition can be transformed into a form in which only the interaction term, the last term, remains. 

On the other hand, for degenerate interaction II, the Coulomb coupling between two atoms can also induce another nonlinear coupling between pairs of modes by tuning the trap frequencies for resonant coupling between the two interacting motional modes. Its Hamiltonian in this case is of the form \cite{Ding2017b},
\begin{align}\label{eq:2}
    \hat{H}_{\rm II}=\hbar\omega_a'\hat{a'}^\dagger\hat{a'}+\hbar\omega_b'\hat{b'}^\dagger\hat{b'}+\hbar\chi\left(\hat{a'}^\dagger\hat{b'}^2+\hat{a'}\left(\hat{b'}^\dagger\right)^2\right).
\end{align}
When the resonance condition $\omega_{a'}=2\omega_{b'}$ is satisfied for degenerate quanta at a frequency $\omega_{b'}$ in this degenerate case, this Hamiltonian is reduced into the last terms in the interaction picture. Here, we distinguish the motional modes of this interaction from that of interaction I by labeling them with the prime symbol. The coupling strength $\chi$ of interaction II is another sensing parameter.  

Noticeably, one may find the interaction Hamiltonians in Eqs. \ref{eq:1} and \ref{eq:2} are rather similar, as replacing $\hat{c}$ and $\hat{c}^\dagger$ with $\hat{b}$ and $\hat{b}^\dagger$ of Eq. \ref{eq:1} turn its form precisely into the form of Eq. \ref{eq:2}. They, however, are intrinsically different and can not be equivalent by straightforwardly relabeling the operator $\hat{c}$ with $\hat{b}$  or vice versa.  In Eq. \ref{eq:1}, $\hat{b}$ and $\hat{c}^\dagger$ commute, while in Eq. \ref{eq:2}, $\hat{b'}$ and $\hat{b'}^\dagger$ do not.

One may recognize that in the strong pumping approximation, the mode operator $\hat{c}$ can be approximated and replaced by its complex pump amplitude $\gamma$. The Hamiltonian in Eq. \ref{eq:1} then resembles the Hamiltonian describing the effect of frequency converters on two frequency modes 
\begin{align}\label{eq:3}
    \hat{H}_{\rm FC}=\hbar\xi\left(\hat{a}^\dagger\hat{b}\gamma+\hat{a}\hat{b}^\dagger\gamma^{*}\right)
\end{align}
where $\xi$ represents the complex conversion parameter, and $\hat{a}$ and $\hat{b}$ are the annihilation operators of the modes entering the input port of the frequency converter. Alternatively, in a strong pumping approximation of mode $a$, its operator can be approximately replaced by its complex amplitude $\hat{a}\rightarrow\alpha$. The Hamiltonian of Eq.\ref{eq:1} then becomes 
\begin{align}\label{eq:3.1}
\hat{H}_{\rm TMS}= \hbar\xi(\alpha\hat{b}\hat{c}+\alpha^{*}\hat{b}\hat{c}),
\end{align}
which is in the form of a two-mode squeezing interaction.
We may also find the similarity between the Hamiltonian in Eq.\ref{eq:2} and the generators of single-mode squeezing operators
\begin{align}\label{eq:4}
    \hat{H}_{\rm Sq}=\frac{{\rm i}\hbar}{2}\left(\zeta \hat{b}^{\dagger 2}+\zeta^*\hat{b}^2\right),
\end{align}
where $\zeta$ is a complex number representing the squeezing parameter, and $\hat{b}$ is the annihilation operator of the single mode. Trilinear interactions in Eqs.\ref{eq:1} and \ref{eq:2} share some similarities to these well-known Hamiltonians of Eqs.\ref{eq:3}, \ref{eq:3.1} and \ref{eq:4}. For example, the operators in Eqs. \ref{eq:1} and \ref{eq:3} both induce the exchange of excitation between different modes of bosons while the operator of Eqs \ref{eq:2} and \ref{eq:4} include a quadratic expression. The trilinear interaction between two modes in Eq.\ref{eq:2}, on the other hand, can also be approximately described by Eq. \ref{eq:4} if mode $a'$ is in a coherent state $\ket{\alpha}$ with $|\alpha|^2\gg 1$. 

Although the trilinear Hamiltonians of Eqs.\ref{eq:1} and \ref{eq:2} can be approximately reduced into these maximally quadratic well-known coupling in some specific circumstances, their intrinsic differences still remain and become explicit if a longer interaction time is considered \cite{Agrawal1974}, or higher efficiency $\xi$ and $\chi$ is available, as in the case of trapped ions and super-conducting circuits. The exact solution for the dynamics of these trilinear interactions can actually be obtained through the iteration method \cite{Carusotto1989}. However, in this work, as the initial motional state is advantageously set to be a Fock state, feasible for both platforms, the dynamics of these two trilinear interactions are more straightforwardly determined, thanks to the bounded dimension of the applicable Hilbert space.

\section{Quantum sensing procedures}\label{sec:protocol}
In this work, we aim to examine if the motional modes, which are treated as a probe, in excited Fock states can enhance the sensitivity of the proposed quantum sensing for estimating the coupling strengths $|\xi|$ and $|\chi|$ of the nonlinear interactions, in Eqs.\ref{eq:1} and \ref{eq:2}. If so, what should be the optimal configuration of the Fock state for a given total number of excitations that provides the highest sensitivity?

\begin{figure}
    \centering
    \includegraphics[width=\textwidth]{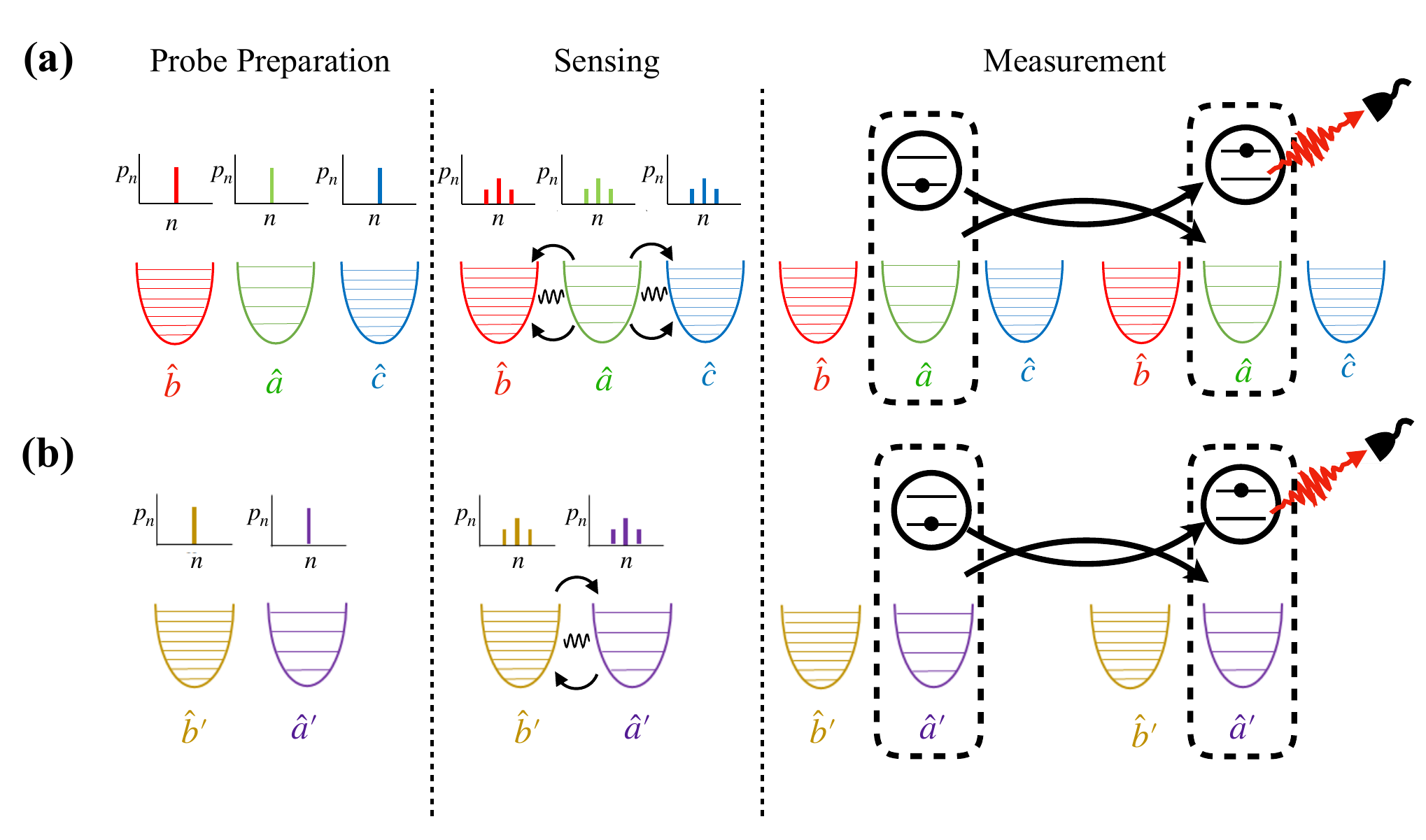}
    \caption{The diagram represents the overall procedure required for sensing the trilinear couplings of {\bf (a)} interaction I and {\bf (b)} interaction II, including three different stages: probe preparation, sensing process and measurement. All available motional modes, including modes $a$ (green), $b$ (red), and $c$ (blue) for interaction I, as well as modes $a'$ (purple) and $b'$ (yellow) for interaction II, are prepared in Fock states and regarded altogether as a probe. We then apply the trilinear coupling between these modes to sense the coupling strength $\xi$ or $\chi$. The motional states thus are altered from their original depending on the value of the strength of the coupling. To monitor the extent of change in the motional states, we couple mode $a$ (green) for interaction I and mode $a'$ (purple) for interaction II with a two-level system via Jaynes-Cummings to perform a sequential measurement of their states. A more challenging technique as explained in \cite{Wolf2019} to measure the motional state in the binary Fock basis, $\{\ket{n}\bra{n}, I - \ket{n}\bra{n}\}$, can also be employed for the measurement.}
    \label{fig:1}
\end{figure}

We propose a quantum scheme as shown in figure \ref{fig:1} for sensing the coupling strengths of both interactions. The scheme includes three standard quantum sensing stages: probe preparation, sensing process and measurement, where the detail of each stage is given as follows. 
The first stage is the probe-preparation stage where the state of each mode is prepared in a Fock state such that the total motional state is initially in a product state of the form
\begin{align}\label{eq:5}
\ket{\psi_0}=\ket{n_a}\otimes\ket{n_b}\otimes\ket{n_c}=\ket{n_a,n_b,n_c},
\end{align}
for sensing the trilinear coupling constant $\xi$ of interaction I, or
\begin{align}\label{eq:6}
    \ket{\psi'_0}=\ket{n_{a'}}\otimes\ket{n_{b'}}=\ket{n_{a'},n_{b'}},
\end{align}
for the case of interaction II. For brevity, from now on, we neglect the tensor product symbol between these motional states and write them in their short form as in the last terms of these equations. For trapped ions, these Fock states can be prepared comprehensively via a standard technique \cite{Leibfried2003, McCormick2019}. We first need to cool these trapped atoms down to their motional ground states $\ket{0}_{i}$, and then sequentially excite them by applying sequences of resonant $\pi$ pulse on their red and blue sidebands so that each $\pi$ pules add one phonon to these modes. This technique achieves a high fidelity with provably increasing quantum non-Gaussianity and force sensing capability \cite{Podhora2022}. Once the probe is satisfactorily prepared, we can perform a nonlinear interaction between these modes to allow them to exchange their excitation for a specific interaction time $t$, causing the probe to change from its original state accordingly. See \cite{Walls1970} for the details of the mathematical method to analytically predict the change of the motional states. The change in its state can be measured through a coupling between the internal electronic states of an atom and the interacting motional modes, allowing us to estimate the coupling strength of the interaction. There are several existing techniques to measure the change of motional Fock states through qubit coupling. However, in this work, we only focus on the measurement using Jaynes-Cummings coupling, as given in \cite{Ritboon2022}, and the binary projective measurement in the Fock basis ${\ket{n}\bra{n}, 1-\ket{n}\bra{n}}$, explained in \cite{Ding2017, Wolf2019}. The projective measurement examines whether the measured mode is in a specific Fock state, $\ket{n}$. The Jaynes-Cummings technique, on the other hand, can be performed sequentially using straightforward experimental techniques, to further retrieve the additional information of the motional state. In the later section, the sensitivity obtained through these techniques is compared and discussed.

After the sensing stage, these motional modes ideally become correlated with each other. Therefore, the choices of motional-mode selection do not affect the final sensitivity for estimating the coupling strength. For example, for non-degenerate interaction I, we can pick any of the three modes to be measured, as ideally, they should give the same sensitivity. For convenience and to prevent confusion in further discussion, modes $a$ and $a'$ are chosen to be the measured modes for interactions I and II respectively as depicted in the figures.

\section{Sensitivity analysis}
We categorize the discussion of the simulated sensitivity obtained from the proposed scheme based on different methods of motional state preparation, including single-mode excitation, two-mode excitation and three-mode excitation. We then investigate each scheme of preparation to find out which configuration provides the highest sensitivity and how the sensitivity depends on the phonon number of the probe. In this work, the sensitivity of the sensing schemes is quantified by its corresponding classical Fisher information.

Here we assume that each stage of the scheme, including motional state preparation, trilinear interactions, and measurement, can be carried out ideally. That means the Fock state of each mode can be prepared perfectly, and the trilinear interactions behave precisely as described in Eqs.\ref{eq:1} and \ref{eq:2} respectively, and the measurement process conforms to its theoretical prediction. We then calculate and analyze the classical Fisher information, defined in the appendix, to evaluate the precision of estimating the coupling strengths, using the probability distribution associated with the measurement outcomes for each preparation scheme. In the simulation results presented below, the interaction times of the nonlinear interactions are both set to be unity.

\subsection{Single-mode excitation}

\subsubsection{Non-degenerate interaction I}
\begin{figure}
    \centering
    \includegraphics[width=\textwidth]{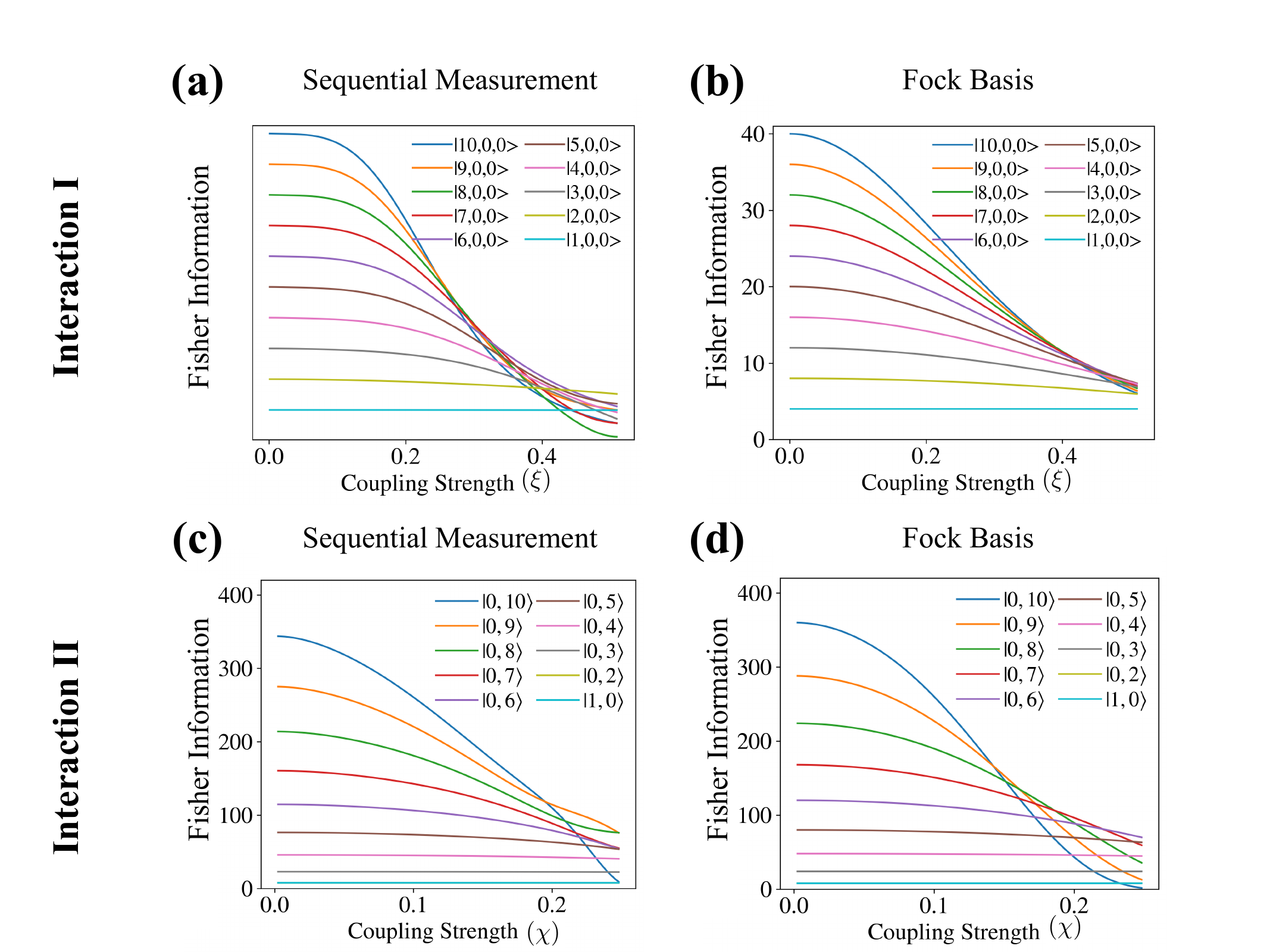}
    \caption{The graphs show the dependence of the Fisher information and the coupling constants of both interactions: $\xi$ ({\bf (a)} and {\bf (b)}) and $\chi$ ({\bf (c)} and {\bf (d)}) for different excitation numbers of 
 the probe.  Only one interacting mode is excited in a Fock state, while the others are set in their ground state. The displayed Fisher information is calculated from the probability distributions associated with a two-shot sequential measurement using JC coupling ({\bf (a)} and {\bf (c)}) and binary Fock basis $\{\ket{n}\bra{n}, I - \ket{n}\bra{n}\}$ ({\bf (b)} and {\bf (d)}). In the regions in which $\xi$ and $\chi$ are very small, the sensitivity increases with the excitation number in exchange for its shortened dynamic ranges.}
    \label{fig:2}
\end{figure}
If we choose to excite only one of the interacting modes and leave the other two in their ground states, the chosen mode must be mode $a$, as interaction I cannot be performed otherwise. To examine the validity of the previous statement, let us assume its contradiction—that modes $a$ and $b$ ($c$) are in their ground state while mode $c$ ($b$) is in a Fock state $\ket{n}_c$ ($\ket{n}_b$). Mode $a$ cannot give its excitation to the other two modes as it is already in its ground state, and it also cannot absorb an excitation from the other modes as, from the Hamiltonian in Eq. \ref{eq:1}, phonons in modes $b$ and $c$ must be absorbed simultaneously, which is impossible as mode $b$ ($c$) is already in its ground state. The interaction is realizable as a result.

Figure \ref{fig:2} shows the simulated results when mode $a$ is excited with different phonon numbers. The classical Fisher information obtained from both measurements, JC and binary projective, increases quadratically with time and linearly with the phonon number of mode $a$ for small values of $\xi$ as
\begin{align}\label{eq:7}
   \lim_{\xi\rightarrow 0} \mathcal{F}^{\rm one}_{\rm {I}}(\xi)=4t^2n_a,
\end{align}
where $\mathcal{F}^{\rm one}_{\rm {I}}$ is the obtained Fisher information and $n_a$ is the excitation number of mode $a$ and $t$ is the trilinear interaction time which is set to be $t=1$.  

On the other hand, the figure also shows that the Fisher information declines as the value of the coupling strength $\xi$ increases, but the decay is not as fast as its first derivative with respect to $\xi$ vanishes. The graphs actually oscillate anharmonically when the coupling strength $\xi$ grows larger than the value specified in the figure. To avoid the unpredictable manner of the Fisher information oscillation, we set the range of effective sensing to be bounded within $\xi_{\rm min}$ such that $0<\xi<\xi_{\rm min}$, where $\xi_{\rm min}$ locates the first local minimum of the Fisher information. The value of $\xi_{\rm min}$, therefore, defines the dynamic range of the sensing. The figure illustrates that $\xi_{\rm min}$ decreases as the phonon number $n_a$ increases and can be approximated by the decreasing rate as
\begin{align}\label{eq:8}
    \xi_{\rm min}\sim\sqrt{\frac{16}{\mathcal{F}^{\rm one}_{\rm I}(0)}}=\sqrt{\frac{4}{n_at^2}},
\end{align}
where $\mathcal{F}^{\rm one}_{\rm I}(0)$ is the Fisher information at $\xi=0$.

\subsubsection{Degenerat interaction II}
For interaction II with a total phonon number $N=1$, a phonon in mode $b'$, the mode with a lower oscillating frequency, is insufficient yet to run the interaction.
However, for $N>1$, the interaction allows either modes $a'$ or $b'$ to be the excited mode. The excitation of mode $b'$ can give higher Fisher information than that of mode $a'$. As shown in figures \ref{fig:2}c and \ref{fig:2}d, the Fisher information is at its maximum near $\chi\sim0$, and related to the phonon number of mode $b'$ as
\begin{align}\label{eq:9}
     \lim_{\chi\rightarrow 0}\mathcal{F}^{\rm one}_{\rm II}(\chi)=4t^2n_{b'}(n_{b'}-1),
\end{align}
for the prepared motional state of the form $\ket{0,n_{b'}}_{a',b'}$, where $n_b'$ is the excitation number of mode $b'$ with mode $a'$ being in its ground state. The Fisher information for sensing small $\chi$ exhibits a quadratic relationship with $n_{b'}$, while it increases linearly with $n_{a'}$. Further details are provided in the following section.

Similar to the case of interaction I, the Fisher information indeed also does not sharply decrease over $\chi$ and oscillates when $\chi$ becomes larger. We then define the upper bound of the dynamic range at which the First local minimum of the Fisher information occurs, says at $\chi=\chi_{\rm min}$. From figures \ref{fig:2}(c) and \ref{fig:2}(d), the values of $\chi_{\rm min}$ decline for higher $n_{b'}$ as
\begin{align}\label{eq:10}
    \chi_{\rm min}\sim\sqrt{\frac{24}{\mathcal{F}^{\rm one}_{\rm II}(0)}}=\sqrt{\frac{6}{n_{b'}(n_{b'}-1)t^2}},
\end{align}
where $\mathcal{F}^{\rm one}_{\rm II}(0)$ is the Fisher information at $\chi =0$.
Comparing the dynamic range in figures \ref{fig:2}(a) and \ref{fig:2}(b) with that in figures \ref{fig:2}(c) and \ref{fig:2}(d), it is apparent that the latter has a shorter dynamic range in general for a given total phonon number. This is reflected by the presence of the quadratic term in the square root of Eq.\ref{eq:10}. We note here that in the figure the Fisher information for $\ket{1,0}$ and $\ket{0,2}$ coincide with each other. 

\subsection{Two-mode excitation}
\subsubsection{Interaction I}
\begin{figure}
    \centering
    \includegraphics[width=\textwidth]{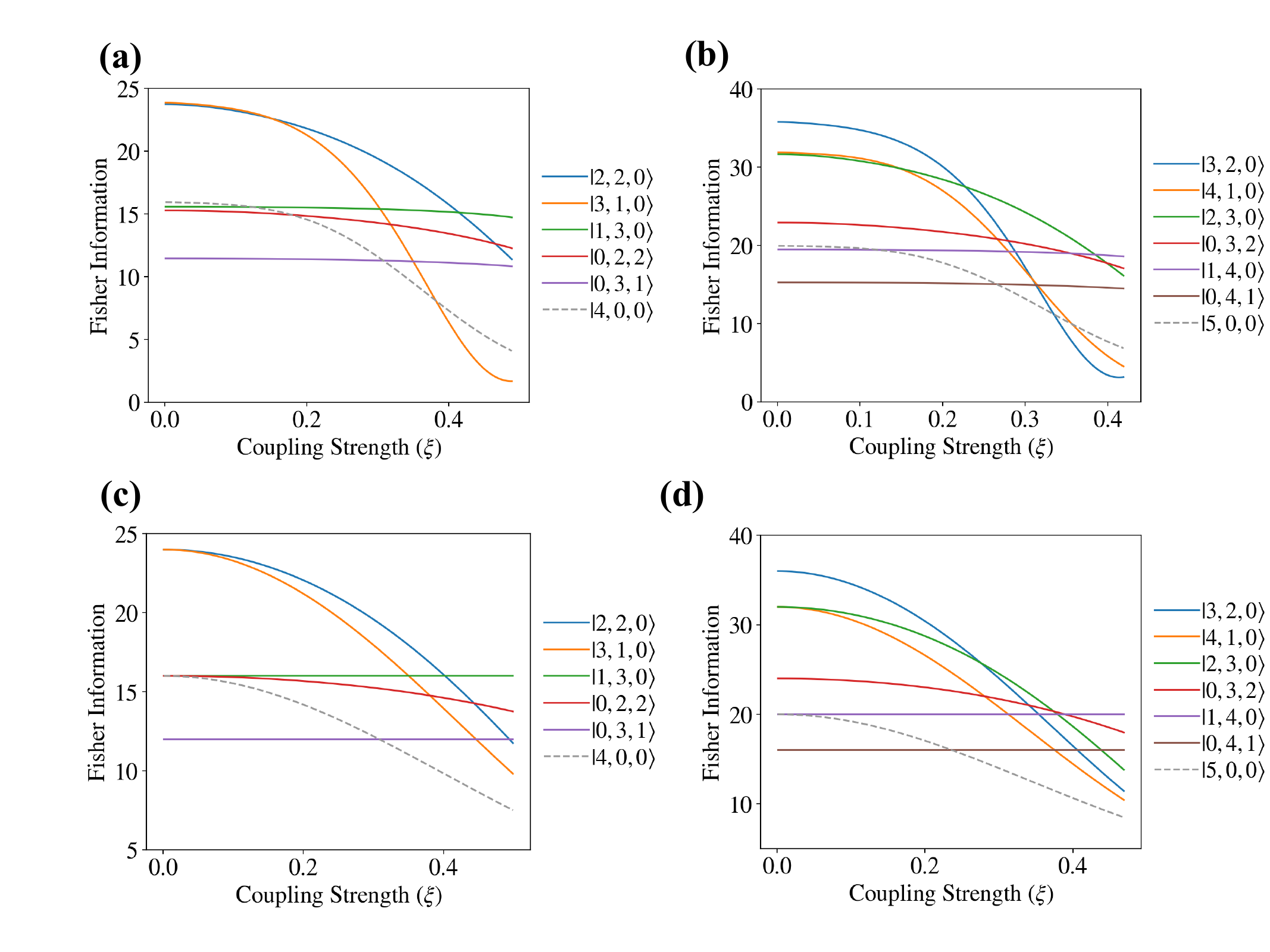}
    \caption{The Fisher information for sensing the coupling strength $\xi$ of interaction I in the two-mode excitation approach with different configurations of probe preparation is displayed. The graphs in figures \ref{fig:3}a and \ref{fig:3}c (figures \ref{fig:3}b and \ref{fig:3}d) show the Fisher information when the total excitation number $N=4$ ($N=5$). We compare the Fisher information obtained from two-shot sequential JC measurements (figures \ref{fig:3}a and \ref{fig:3}b) with that obtained from the measurement in the projective Fock basis ${\ket{n}\bra{n}, I - \ket{n}\bra{n}}$ (figures \ref{fig:3}c and \ref{fig:3}d). The configuration of the probe satisfying Eqs.\ref{eq:13} and \ref{eq:14} yields optimal sensitivity, which is evidently higher than that obtained from the single-mode excitation approach, as indicated by the dashed grey lines.}
    \label{fig:3}
\end{figure}
We now shift our interest to the case of two-mode excitation, where only two of the interacting modes are chosen to be excited, and determine the optimal configuration of Fock states given the best sensitivity. In the case of interaction I, this excitation approach leaves one of the three modes in its ground state. If we consider the sensitivity analysis, given in the appendix, the Fisher information for a small value of $\xi$ can be written as
\begin{align}\label{eq:11}
\lim_{\xi\rightarrow 0}\mathcal{F}^{\rm two}_{\rm I}(\xi)=4t^2(n_an_{b(c)}+n_{a}),
\end{align}
when either mode $b$ or $c$ is in its ground state while the others in Fock states, or
\begin{align}\label{eq:12}
    \lim_{\xi\rightarrow 0}\mathcal{F}^{\rm two}_{\rm I}(\xi)=4t^2n_b n_c,
\end{align}
if mode $a$ is chosen to be in the ground state instead, where $n_i$ is a phonon number of mode $i$. Instead of limiting the overall required excitation energy for estimation, we operationally limit the overall number of quanta required to prepare the Fock states, as it corresponds to the number of $\pi$-pulses needed in the state preparation. It is apparent that for a given total phonon number $N=n_{a}+n_{b(c)}$ the former case should give us better sensitivity due to the second term of Eq.\ref{eq:11}. This fact is shown clearly in figure \ref{fig:3}, as greater Fisher information is achieved if mode $a$ together with either mode $b$ or $c$ is excited. We use the Lagrange multiplier method to estimate the optimal configuration of Fock states for this excitation approach.  Using the Lagrange multiplier method, we estimate the optimal configuration of Fock states for this excitation approach. We find that for a given phonon number $N$, such an optimal state should have the numbers of phonons in mode $a$ and the other chosen mode to be
\begin{align}\label{eq:13}
    n_a=\frac{N+1}{2},\quad n_{b(c)} =\frac{N-1}{2},
\end{align}
if $N$ is odd and
\begin{align}\label{eq:14}
    n_a= n_{b(c)}=\frac{N}{2},
\end{align}
if $N$ is even.
Due to the second term of Eq.\ref{eq:11}, to achieve greater Fisher information, mode $a$ is prioritized to have a greater number of excitations than the other mode. Lastly, compared to the dashed grey line in figure \ref{fig:3}, we can clearly see that two-mode excitation can give greater sensitivity than the single-mode excitation approach.

Similar to the previous case, the higher Fisher information at $\xi\sim 0$ results in a shorter dynamic range,
\begin{align}\label{eq:15}
\xi_{\rm min}\sim\sqrt{\frac{16}{\mathcal{F}^{\rm two}_{\rm I}(0)}},
\end{align}
however, with an equivalent prefactor to the case of Eq.\ref{eq:8} in the single-mode excitation scheme. Remarkably, some configurations, such as $|1,3,0\rangle$ or $|0,3,1\rangle$ in figure \ref{fig:4}a, with lower Fisher information, exhibit stable and large dynamical ranges. This aspect, similar to other panels of figure \ref{fig:4}, shows an apparent and strong trade-off between the maximum Fisher information and the dynamic range of estimation.

\subsubsection{Interaction II}
\begin{figure}
    \centering
    \includegraphics[width=\textwidth]{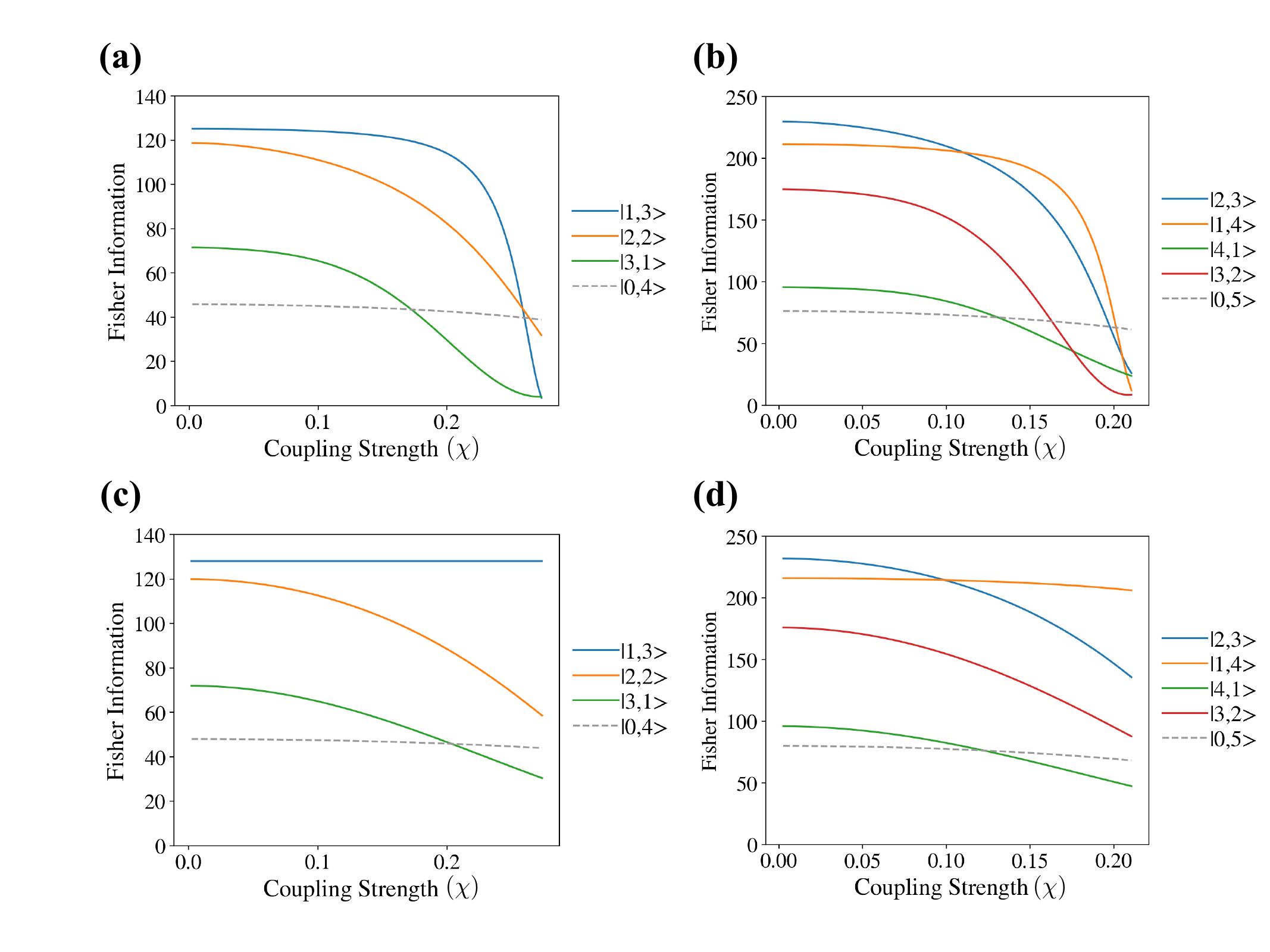}
    \caption{The Fisher information for sensing the coupling strength $\chi$ of interaction II is presented for various configurations where both interacting motional modes are excited. The graphs in figures \ref{fig:4}a and \ref{fig:4}c (figures \ref{fig:4}b and \ref{fig:4}d) display the Fisher information when the total phonon number $N=4$ ($N=5$). The Fisher information of different configurations using two-shot sequential JC measurements, in figures \ref{fig:4}a and \ref{fig:4}b, and the measurement in the projective Fock basis ${\ket{n}\bra{n}, I - \ket{n}\bra{n}}$, in figures \ref{fig:4}c and \ref{fig:4}d, is compared. The sensitivity is obviously higher than that given by single-mode excitation, displayed by dashed grey lines. The optimal configuration is achieved when the probe is prepared in a state where mode $b'$ has approximately twice as much excitation as mode $a'$, reflecting the ratio between $\hat{a}'$ and $\hat{b}'$ in the interaction Hamiltonian of Eq.\ref{eq:2}.}
    \label{fig:4}
\end{figure}
In interaction II, both interacting modes are excited, represented by the Fock state $\ket{n_{a'},n_{b'}}{a',b'}$. The sensitivity analysis of the sensing, see the appendix, informs us how the Fisher information for small coupling strength $\chi\ll 1$, depends on the excitation number of each mode as
\begin{align}\label{eq:16}
\lim{\chi\rightarrow 0}\mathcal{F}{\rm II}(\chi)=4t^2\left[n{b'}(n_{b'}-1)(n_{a'}+1)+(n_{b'}+1)(n_{b'}+2)n_{a'}\right].
\end{align}
It is evident that the Fisher information quadratically increases with the number of phonons in mode $b'$. The probe, therefore, becomes more sensitive if we excite the motion in this mode more than the other, which can only provide a linear increase, see figure \ref{fig:4}. We can employ the Lagrange multiplier method, as before, to estimate the optimal configuration. Even though the prediction cannot be exactly given by the method, we can still roughly infer the pattern of the optimal Fock states. Depending on a given total phonon number $N=n_{a'}+n_{b'}$, the optimal configuration becomes
\begin{align}\label{eq:17}
\ket{n_{a'},n_{b'}}=\ket{n, 2n};\ n=\frac{N}{3},
\end{align}
when $N$ is a multiple of 3, or
\begin{align}\label{eq:18}
\ket{n_{a'},n_{b'}}=\ket{n, 2n+1};\ n=\frac{N-1}{3}
\end{align}
for $N\mod 3 = 1$, or, lastly,
\begin{align}\label{eq:19}
\ket{n_{a'},n_{b'}}=\ket{n+1, 2n+1};\ n=\frac{N-2}{3},
\end{align}
for $N \mod 3 = 2$. For example, if the total phonon number is 4, where $4\mod 3 = 1$, using Eq. \ref{eq:18}, the optimal Fock state, in this case, becomes $\ket{1,3}_{a',b'}$ as shown in figures \ref{fig:4}a and \ref{fig:4}c.

Eqs.\ \ref{eq:17} to \ref{eq:19} indicate how the excitation is optimally shared between the two modes. Roughly two-thirds of the excitation should be given to mode $b'$, with the remaining third to mode $a'$. We may thus roughly say that for a large number $N\gg 1$, the best possible Fisher information of this scheme for sensing around $\chi\ll 1$ becomes proportional to $N^3$ as
\begin{align}\label{eq:20}
\lim_{\chi\rightarrow 0}\mathcal{F}_{\rm II}(\chi)\sim \frac{32N^3}{27}t^2.
\end{align}

The dynamic range again declines with the gained Fisher information at $\chi=0$, which can be estimated by
\begin{align}\label{eq:21}
\chi_{\rm min}\sim \sqrt{\frac{24}{\mathcal{F}{\rm II}(0)}}.
\end{align}
Using the optimal configuration, its relation with the total excitation number $N$ can be roughly expressed as
\begin{align}\label{eq:22}
\chi{\rm min}\sim \sqrt{\frac{81}{4N^3t^2}}.
\end{align}
From the figure, and Eqs.\ \ref{eq:11} and \ref{eq:20}, it is apparent that two-mode excitation can give better sensitivity for small $\chi$ compared to the single-mode excitation approach.

\subsection{Three-mode excitation}
\begin{figure}
    \centering
    \includegraphics[width=\textwidth]{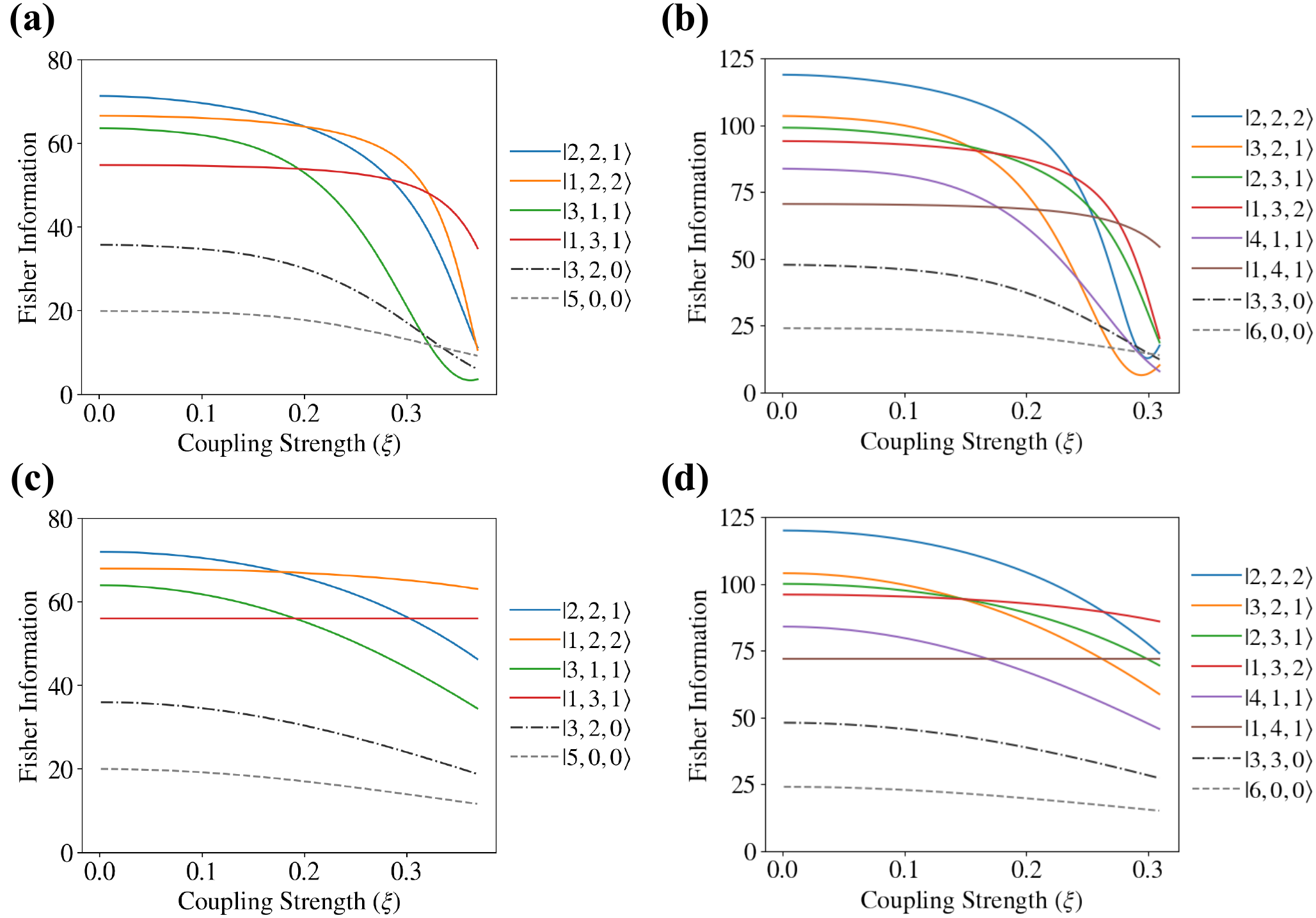}
    \caption{The Fisher information for sensing the coupling strength $\xi$ of interaction I using the three-mode excitation approach with different configurations of probe preparation. The first and second columns of the graphs display the Fisher information when the total phonon numbers are ({\bf a} and {\bf c}) 5 and ({\bf b} and {\bf d}) 6 respectively. The Fisher information of different configurations using from two-shot sequential JC measurements, in figures \ref{fig:5}a and \ref{fig:5}b, and the measurement in the projective Fock basis $\{\ket{n}\bra{n}, I - \ket{n}\bra{n}\}$, in figures \ref{fig:5}c and \ref{fig:5}d, is compared. The optimal sensitivity for small $\xi$ can be achieved if the initial excitation quanta of the interacting modes shares as evenly as possible reflecting the ratio $1:1:1$ of the mode operators $\hat{a}$, $\hat{b}$ and $\hat{c}$ in the interaction Hamiltonian,  Eq.\ref{eq:1}. The optimal configurations of this state preparation approach give higher Fisher information than the single-mode (dashed grey line) and two-mode (dash-dotted black line) excitation approaches.}
    \label{fig:5}
\end{figure}
We then consider the case when three of the modes are excited in Fock states. It is only interaction I that has enough motional modes to satisfy this condition. From the appendix, the Fisher information, for sensing $\xi\ll 1$, is related to the phonon numbers of all interacting modes as
\begin{align}\label{eq:23}
    \lim_{\xi\rightarrow 0}\mathcal{F}^{\rm max}_{\rm I}(\xi)=4t^2\left[n_a(n_b+1)(n_c+1)+(n_a+1)n_bn_c\right],
\end{align}
where $n_i$ is the phonon number of mode $i$. It is obvious from the formula that for interaction I, this excitation approach can give  Fisher information higher than the previous two approaches, the single-mode and two-mode excitation, for a given total phonon number $N=n_a+n_b+n_c$. Lagrange multiplier is again employed to estimate the optimal configuration of Fock states. It turns out that for a given  $N$, the optimal Fock states are the states whose excitations in each mode are shared approximately evenly where mode $a$ is slightly more prioritized than the others. This is due to the interaction Hamiltonian in Eq. \ref{eq:1} has the same number of annihilation or creation operators for these three modes. To be precisely specific, let us consider the following cases of total phonon numbers $N$. For $N \mod 3 =0$,  the excitation in each mode must be shared evenly, {\it i.e.},
\begin{align}\label{eq:24}
\ket{n_a,n_b,n_c}=\ket{n,n,n}_{a,b,c},
\end{align}
where $n=N/3$, see figure \ref{fig:5}b and \ref{fig:5}d for $N=6$. For the case of $N\mod 3\neq 0$, we prioritize mode $a$ to have more excitations than the other modes. For example, for $N\mod 3=1 $, the optimal Fock state is thus of the form
\begin{align}\label{eq:25}
\ket{n_a,n_b,n_c}=\ket{n+1,n,n}_{a,b,c},
\end{align}
where $n=(N-1)/3$. Finally, for $N\mod 3 =2$, the optimal Fock state  can either be  
\begin{align}\label{eq:26}
\ket{n_a,n_b,n_c}=\ket{n+1,n+1,n}_{a,b,c}\ \ \ \text{or}\ \ \  \ket{n_a,n_b,n_c}=\ket{n+1,n,n+1}_{a,b,c}
\end{align}
where $n=(N-2)/3$, as these two states give the same optimal Fisher information due to the Hamiltonian symmetry of modes $b$ and $c$. For example, as shown in figure \ref{fig:5}a and \ref{fig:5}c for $N=5$, the optimal state is obviously $\ket{2,2,1}_{a,b,c}$. As the excitation of the optimal Fock states is shared evenly in each mode, it indicates that for a large total phonon number $N\gg1$ the optimal Fisher information for $\xi\ll 1$ is approximately
\begin{align}\label{eq:27}
    \lim_{\xi\rightarrow 0}\mathcal{F}_{\rm I}(\xi)\sim \frac{8t^2N^3}{27}.
\end{align}
The proportionality of the Fisher information to $N^3$ resembles the case of two-mode excitation for interaction II. This is due to the similarity of their interaction Hamiltonians.

Similar to the previous case, the Fisher information is at its highest around $\xi \sim 0$ and decreases as the coupling strength grows larger. The dynamic range, again, becomes shortened for larger Fisher information $\mathcal{F}_{\rm I}$ at $\xi \ll 1$ as
\begin{align}\label{eq:28}
    \chi_{\rm min}\sim \sqrt{\frac{24}{\mathcal{F}_{\rm I}(0)}}.
\end{align}
If the probe is in optimal states, the dynamic range thus approximately becomes
\begin{align}\label{eq:29}
    \xi_{\rm min}\sim\sqrt{\frac{81}{N^3t^2}}.
\end{align}

\section{Discussion}
The result from the last section provides us with the optimal configurations for each proposed excitation approach. It's evident that the optimal Fisher information is achieved when all motional modes are excited in Fock states. For interaction I, the three-mode excitation approach offers the highest sensitivity compared to the other considered excitation approaches.

In the case of interaction I, with optimal configurations, the Fisher information is proportional to $N^3$, as predicted by Eq. \ref{eq:27}. The difference between the solid lines and the dashed lines in Figure \ref{fig:1}a highlights the significance of the next-to-leading order term, which is proportional to $N^2$. The single- and two-mode excitation approaches, however, exhibit linear and quadratic relationships, respectively, as predicted earlier. This suggests that the probe is most sensitive to the trilinear coupling when all interacting modes are nearly evenly excited reflecting the ratio $1:1:1$ of the three interacting mode operators in the interaction Hamiltonian.

Similarly, for interaction II, when all two modes are strategically excited as indicated by Eqs. \ref{eq:17}-\ref{eq:19}, the sensitivity increases cubically with the excitation number $N$. However, their relation reduces to quadratic using the single-mode excitation approach. Moreover, for a given total excitation $N$, the Fisher information for sensing interaction II is noticeably higher than that of interaction I because its interaction Hamiltonian contains quadratic terms of annihilation and creation operators of an interacting mode.

One may also notice that particular Fock states can yield a very large dynamic range of sensitivity for sensing, such as the states $\ket{0,3,1}$ and $\ket{1,4,0}$ in figures \ref{fig:3}c and \ref{fig:3}d, respectively. This is because their time evolution can be expressed within a sufficiently small and finite Hilbert space during the sensing process. As a result, a measurement on an interacting motional mode can provide the full information of the probe. This fact also holds true for interaction II, as the state $\ket{1,3}$ in figure \ref{fig:4}c also exhibits a very large dynamic range.

We emphasize here that preparing Fock states in trapped ions is rather pragmatic with the already existing experimental techniques and technologies compared to other candidate states. Moreover, as shown in the appendix, the sensitivity of the proposed protocol does not completely collapse if the probe is not prepared perfectly in an ideal Fock state.

\section{Conclusion}
We propose a sensing protocol using the probe in an available Fock state to measure the coupling strength of nonlinear interactions, which are crucial for various fundamental tests and applications. The probe, when prepared in Fock states, proves to be optimal for sensing such nonlinear interactions, as demonstrated using interactions I and II as examples. The high sensitivity of a Fock state $\ket{n}$ arises from its distinct population distribution. In the ideal scenario, only $\ket{n}$ has population, and any small disturbance from the interaction induces a change in the population distribution. This change can technically be measured by coupling the probe with a qubit during the measurement stage. The classical Fisher information, which quantifies the sensitivity of the protocol, is calculated using two different types of measurements: sequential measurement using JC coupling and projective measurement using the Fock basis.

 We found that the best optimal configuration of Fock state for sensing of the non-degenerate and degenerate interactions I and II strongly depends on the ratio of the mode operators in the interaction Hamiltonian. In other words, the optimal configuration for sensing a nonlinear interaction depends on how the motional modes interact with each other. In such configuration the Fisher information increases cubically with the total excitation quanta $N$, reflecting the estimation errors scaled down by $N^{-3}$. 
 
 The high sensitivity provided by a high quality of Fock state can help us reveal unexplored aspects of the dynamics of nonlinear interactions in bosonic experiments including trapped ions and superconducting circuits bringing further development in quantum technology and processing.

\begin{acknowledgements}
We acknowledge the support of the project CZ.02.01.01/00/22\_008/0004649 of the Czech Ministry of Education, Youth and Sport, 22-27431S of Czech Science Foundation and Thammasat University Research Unit in Energy Innovations and Modern Physics (EIMP). A.R. acknowledges funding support from the NSRF via the Program Management Unit for Human Resources \& Institutional Development, Research and Innovation (grant number B39G670018) and the Thailand Science Research and Innovation Fundamental Fund fiscal year 2024, Thammasat University. R.F. acknowledges funding from the MEYS of the Czech Republic (Grant Agreement 8C22001). Project SPARQL has received funding from the European Union’s Horizon 2020 Research and Innovation Programme under Grant Agreement No. 731473 and 101017733 (QuantERA). 
\end{acknowledgements}

\appendix
\section{Classical Fisher information}
The classical Fisher information quantifies the sensitivity of a measured observable $x$ with respect to the change of a sensing parameter $\theta$, which is defined as \cite{Paris2009,Cramer1999,Rao1992}
\begin{align}\label{eq:a1.1}
\mathcal{F}(\theta)= \sum_{x}\frac{1}{P(x|\theta)}\left(\frac{\partial P(x|\theta)}{\partial \theta}\right)^2,
\end{align}
where $P(x|\theta)$ is the probability distribution of the measured observable $x$ when the sensing parameter is of value $\theta$. The probability distribution can be obtained through a quantum measurement of the observable $x$ as $P(x|\theta)={\rm Tr}(\hat{\Pi}_{x}\rho(\theta))$, where $\rho(\theta)$ is the quantum state after performing the sensing with the sensing parameter $\theta$. The Fisher information can also be used to determine the bound of the precision of the estimation as
\begin{align}\label{eq:a1.2}
\Delta\theta_{\rm est} \geq \Delta_{\rm CR}=\frac{1}{\sqrt{N\mathcal{F}(\theta)}},
\end{align}
where $\Delta\theta_{\rm est}$ is the standard deviation of an estimator $\theta_{\rm est}$ of the sensing parameter $\theta$, and $\Delta\theta_{\rm CR}$ is the Cram{\'e}r-Rao bound and $N$ is the number of trails in the measurement.

\section{Quantum Fisher information}\label{sec:ap.2}
The Quantum Fisher information (QFI) can be obtained through maximizing the Fisher information $\mathcal{F}(\theta)$ over all possible quantum measurements\cite{Paris2009}, 
\begin{align}\label{eq:a2.1}
\max_{\{\hat{\Pi}_x\}}\mathcal{F}(\theta)=\mathcal{F}_{\rm Q}[\rho_0,\hat{G}],
\end{align}
which depends only on the prepared initial state $\rho_0$ and the generator $\hat{G}$ of the unitary evolution of the sensing. Therefore, QFI allows us to quantify the lower bound of the estimation error for a given initial state and a unitary evolution as
\begin{align}\label{eq:a2.2}
\Delta\theta_{\rm est}\geq\Delta\theta_{\rm CR}\geq\Delta\theta_{\rm QCR}=\frac{1}{\sqrt{N\mathcal{F}_{\rm Q}}},
\end{align}
where $\Delta\theta_{\rm QCR}$ is the quantum Cram{\'e}r-Rao bound which is inversely proportional to the square root of the trail number $N$ and the QFI.
For an initial pure state $\rho_0=\ket{\psi_0}\bra{\psi_0}$, the QFI can be determined simply by the variance of the generator $\hat{G}$, as described in \cite{Paris2009}, as
\begin{align}\label{eq:a2.3}
    \mathcal{F}_{\rm Q} [\ket{\psi_0},\hat{G}] = 4\bra{\psi_0}\Delta\hat{G}^2\ket{\psi_0}, 
\end{align}
where $\Delta\hat{G}^2=\hat{G}^2-\left|\bra{\psi_0}\hat{G}\ket{\psi_0}\right|^2$. 

As we consider the interaction pictures under the resonance condition, the Hamiltonians in Eqs. \ref{eq:1} and \ref{eq:2} are reduced to
\begin{align}\label{eq:a2.4}
\hat{H}^{\rm i}_{\rm I}=\hbar \xi\hat{G}_{\rm I},
\end{align}
with $\hat{G}_{\rm I}=(\hat{a}^\dagger\hat{b}\hat{c}+\hat{a}\hat{b}^\dagger\hat{c}^\dagger)$ for interaction I, and
\begin{align}\label{eq:a2.5}
\hat{H}^{\rm i}_{\rm II}=\hbar \chi \hat{G}_{\rm II},
\end{align}
with $\hat{G}_{\rm II}=\left(\hat{a}^\dagger\hat{b}^2+\hat{a}\left(\hat{b}^\dagger\right)^2\right)$ for interaction II. Treating $\xi$ and $\chi$ as the sensing parameters to be estimated, the generators of the time evolution of these interactions are proportional to $t\hat{G}{\rm I,II}$ as their unitary evolution is of the form,
\begin{align}\label{eq:a2.6}
\hat{U}_{\rm I,II}(t)=\exp(-{\rm i}\hat{H}_{\rm I,II}t/\hbar)=\exp(-{\rm i}\lambda_{\rm I, II}\hat{G}_{\rm I,II} t),
\end{align}
where $\lambda_{\rm I}=\xi$ and $\lambda_{\rm II}=\chi$  are the couplings of the interactions.
This allows us to straightforwardly quantify the QFI of these interactions when an initial state is a tensor product of Fock states by evaluating the variance of their generators $\hat{G}_{\rm I}$ and $\hat{G}_{\rm II}$. The QFI for estimating the coupling strength $\xi$ of interaction I for a tensor product state $\ket{n_a,n_b,n_c}$ is given as
\begin{align}\label{eq:a2.7}
\mathcal{F}_{\rm Q, I}=4t^2\left(n_a(n_b+1)(n_c+1)+(n_a+1)n_bn_c\right).
\end{align}
In the same manner, for the probe in the product Fock state $\ket{n_{a'},n_{b'}}$, the QFI for the interaction II can be obtained as
\begin{align}\label{eq:a2.8}
    \mathcal{F}_{\rm Q, II}=4t^2\left(n_{b'}(n_{b'}-1)(n_{a'}+1)+(n_{b'}+1)(n_{b'}+2)n_{a'}\right).
\end{align}
These two equations display the excitation dependence of these QFIs, reflecting the cubic increase in the probe's sensitivity due to the increase in the total number of excitations $N$.
\section{Optimal configurations}
The optimal configuration of the Fock states that gives the highest Quantum Fisher information for interactions I and II can be obtained through the help of the Lagrange multiplier. Even though the excitation number $n_i$ of mode $i$ in Eqs.\ref{eq:a2.7} and \ref{eq:a2.8} is, in fact, natural numbers, we, at this point, treat it as a continuous variable, so that we can find the value of each $n_i$ that maximize the Quantum Fisher information for a given total excitation $N$.

In the case of interaction I, we optimize the value of the quantum Fisher information $\mathcal{F}_{\rm Q, I}$ of Eq.\ref{eq:a2.7} by finding the optimal values of $n_a$, $n_b$ and $n_c$, given that $N=n_a+n_b+n_c$ is the total excitation of the prepared state. The equation constraint for the Lagrange multiplier technique thus becomes $\mathcal{C}(n_a,n_b,n_c)=n_a+n_b+n_c-N$. To determine each optimal $n_i$, we, therefore, have to numerically find  the roots of the following equation system,
\begin{align}
\frac{\partial}{\partial n_a}\mathcal{F}_{\rm Q, I}&=\lambda \mathcal{C}(n_a,n_b,n_c),\\
\frac{\partial}{\partial n_b}\mathcal{F}_{\rm Q, I}&=\lambda \mathcal{C}(n_a,n_b,n_c),\\
\frac{\partial}{\partial n_c}\mathcal{F}_{\rm Q, I}&=\lambda \mathcal{C}(n_a,n_b,n_c),\\
N&=n_a+n_b+n_c.
\end{align}
The values of $n_a$, $n_b$ and $n_c$ which satisfy these equations, of course, are not integers. For example, in the case of $N=n_a+n_b+n_c=4$, we numerically find $n_a=1.52$, $n_b=n_c=1.24$. However, we can easily realize that the closed integers of these three numbers should be $n_a=2$, $n_b=n_c=1$ giving the optimal configuration for $N=4$ to be the Fock state $\ket{2,1,1}$. The optimal configuration of Fock states for interaction II is determined in the same manner. We thus can find the pattern of the optimal configuration for a given excitation as discussed in the main text. We also note that the method can be also used to find the optimal configuration for the probes when constraint to overall energy is used.

\section{Quantum state disturbance for short-time interaction}
\subsection{Interaction I}
The short-time evolution of the probe state initially prepared in a Fock state $\ket{n_a,n_b,n_c}$ can be determined by the polynomial expansion of the unitary operator given in Eq. \ref{eq:a2.6} as
\begin{align}\label{eq:a3.1}
\ket{\psi(\delta t)}&=\hat{U}_{\rm I}(\delta t)\ket{n_a,n_b,n_c}\nonumber\\
&=\exp(-{\rm i}\hat{H}_{\rm I}\delta t/\hbar)\ket{n_a,n_b,n_c}\nonumber\\
&=\left(I-{\rm i}\delta t\xi(\hat{a}^\dagger\hat{b}\hat{c}+\hat{a}\hat{b}^\dagger\hat{c}^\dagger)-\frac{\delta t^2}{2!}\left(\hat{a}^\dagger\hat{b}\hat{c}+\hat{a}\hat{b}^\dagger\hat{c}^\dagger\right)^2+\mathcal{O}(\delta t^3)\right)\ket{n_a,n_b,n_c}\nonumber\\
&\approx \ket{n_a,n_b,n_c}-{\rm i}\delta t\xi\sqrt{n_a(n_b+1)(n_c+1)}\ket{n_a-1,n_b+1,n_c+1}\nonumber\\
& \quad \quad -{\rm i}\delta t\xi\sqrt{(n_a+1)n_b n_c)}\ket{n_a+1,n_b-1,n_c-1}, 
\end{align}
where $\delta t$ represents a small interaction time such that $\delta t\ll 1$. 
The approximated reduced state of mode $a$ at a small evolution time $\delta t$ can thus be expressed as
\begin{align}\label{eq:a3.2}
\rho_a &= \Tr_{b,c}\left(\ket{\psi(\delta t)}\bra{\psi(\delta t)}\right)\nonumber\\
&\approx \ket{n_a}\bra{n_a}+\delta t^2\xi^2n_a(n_b+1)(n_c+1)\ket{n_a-1}\bra{n_a-1} \nonumber\\
& \quad \quad +\delta t^2\xi^2(n_a+1)n_b n_c\ket{n_a+1}\bra{n_a+1}.
\end{align}
We note here that the deviation from the initial state Fock state $\ket{n_a}$ of mode $a$ is proportional to the multiplication of the excitation numbers of the interacting motional modes as mentioned in the main text.
The approximate population distribution after the sensing process becomes
\begin{align}
p_{n_a} &\approx 1-\delta t^2\xi^2\left(n_a(n_b+1)(n_c+1)+(n_a+1)n_b n_c\right)\label{eq:a3.3}\\
p_{n_a+1} &\approx \delta t^2\xi^2 n_a(n_b+1)(n_c+1)\label{eq:a3.4}\\
p_{n_a-1} &\approx \delta t^2\xi^2 (n_a+1)n_b n_c.\label{eq:a3.5}
\end{align}
By substituting these probabilities for small $\xi$ distribution in Eq. \ref{eq:a1.1}, we find
\begin{align}\label{eq:a3.6}
    \mathcal{F}(\xi) &\approx \frac{1}{p_{n_a}}\left(\frac{\partial p_{n_a}}{\partial \xi}\right)^2+\frac{1}{p_{n_a+1}}\left(\frac{\partial p_{n_a+1}}{\partial \xi}\right)^2+\frac{1}{p_{n_a-1}}\left(\frac{\partial p_{n_a-1}}{\partial \xi}\right)^2 \nonumber\\
    &\approx 0+ 4\delta t^2 n_a(n_b+1)(n_c+1)+ 4\delta t^2 (n_a+1)n_b n_c, \nonumber\\
    &=4\delta t^2\left(n_a(n_b+1)(n_c+1)+(n_a+1)n_b n_c\right),
\end{align}
where the first term is neglected as it is much smaller than the remaining.
In fact, for small $\xi$, any measurement that can discriminate the state $\ket{n_a}$ from its two neighboring states $\ket{n_a+1}$ and $\ket{n_a-1}$ can clearly give Fisher information reaching the QFI of Eq.\ref{eq:a2.7}, as, from Eqs. \ref{eq:a3.3}-\ref{eq:a3.5}, we can simply show that
\begin{align}
    \frac{1}{p_{n_a+1}}\left(\frac{\partial p_{n_a+1}}{\partial \xi}\right)^2+\frac{1}{p_{n_a-1}}\left(\frac{\partial p_{n_a-1}}{\partial \xi}\right)^2=\frac{1}{p_{n_a+1}+p_{n_a-1}}\left(\frac{\partial}{\partial \xi}(p_{n_a+1}+p_{n_a-1})\right)^2.
\end{align}

\subsection{Interaction II}

For interaction II, we can calculate the short-time disturbance of the prepared Fock state in the same manner as the case of interaction I. The short-time evolution of the initial state $\ket{n_{a'},n_{b'}}$ is given as
\begin{align}
\ket{\psi(\delta t)}&=\hat{U}_{\rm II}(\delta t)\ket{n_{a'},n_{b'}}\nonumber\\
&=\exp(-{\rm i}\hat{H}_{\rm II}\delta t/\hbar)\ket{n_{a'},n_{b'}}\nonumber\\
&=\left(I-{\rm i}\delta t\chi\left(\hat{a'}^\dagger\hat{b'}^2+\hat{a}\left(\hat{b'}^\dagger\right)^2\right)+\mathcal{O}(\delta t^2)\right)\ket{n_{a'},n_{b'}}\nonumber\\
&\approx \ket{n_{a'},n_{b'}}-{\rm i}\delta t\chi\sqrt{n_{b}(n_{b'}+1)(n_{a'}+1)}\ket{n_{a'}+1,n_{b}-2}\nonumber\\
& \quad \quad -{\rm i}\delta t\chi\sqrt{(n_{b'}+1)(n_{b'}+2)n_{a'}}\ket{n_{a'}-1,n_{b'}+2}, 
\end{align}
with again $\delta t\ll 1$. The state of mode $a'$ in this time regime is thus given as
\begin{align}
   \rho_{a'} &\approx \ket{n_{a'}}\bra{n_{a'}}+\delta t^2\chi^2(n_{b'}+1)(n_{b'}+2)n_{a'}\ket{n_{a'}-1}\bra{n_{a'}-1} \nonumber\\
& \quad \quad +\delta t^2\chi^2 n_{b}(n_{b'}+1)(n_{a'}+1)\ket{n_{a'}+1}\bra{n_{a'}+1}.
\end{align}
We thus can find the probability distribution and use it to determine the classical Fisher information. The Fisher information for short interaction time is given as
\begin{align}
    \mathcal F({\chi})&\approx\frac{1}{p_{n_{a'}}}\left(\frac{\partial p_{n_{a'}}}{\partial \chi}\right)^2+\frac{1}{p_{n_{a'}+1}}\left(\frac{\partial p_{n_{a'}+1}}{\partial \chi}\right)^2+\frac{1}{p_{n_{a'}-1}}\left(\frac{\partial p_{n_{a'}-1}}{\partial \chi}\right)^2\nonumber\\
    &\approx 4\delta t^2\left((n_{b'}+1)(n_{b'}+2)n_{a'}+n_{b}(n_{b'}+1)(n_{a'}+1)\right),
\end{align}
which saturates the QFI for small $\chi$.

\section{Sequential phonon measurement using qubit}
As reported in \cite{Ritboon2022}, sequential measurements on phonons using the internal states of atoms as an ancilla qubit can give more information about the phononic state compared to the traditional single-shot measurement.It is especially useful for measuring small disturbances in Fock states. The measurement outcome can distinguish whether the phononic state remains in the original prepared state $\ket{n}$ or becomes disturbed, transitioning to the neighboring states $\ket{n+1}$ or $\ket{n-1}$, or even to the next-to-neighboring states $\ket{n+2}$ or $\ket{n-2}$. This scheme of sequential measurements is approximately equivalent to a von Neumann projective measurement of phonons in the basis
\begin{align}
S_0=\left\{\hat{\Pi}_0, \hat{\Pi}_1, \hat{\Pi}_2, \hat{\Pi}_3\right\},    
\end{align}
with
\begin{align}
    \hat{\Pi}_0&=\ket{n}\bra{n},\\
    \hat{\Pi}_1&=\ket{n-1}\bra{n-1}+\ket{n+1}\bra{n+1},\\
    \hat{\Pi}_2&=\ket{n-2}\bra{n-2}+\ket{n+2}\bra{n+2},\\
    \hat{\Pi}_3&=I-\sum^{2}_{i=0}\Pi_i,
\end{align}
where $I$ represents the identity in the motional vector space, see more details in \cite{Ritboon2022}. The additional information from this scheme provides a lower decay rate of Fisher information at the coupling strength near zero compared to the decay rate obtained from measurements made in the projective Fock basis $\{\ket{n}\bra{n},1-\ket{n}\bra{n}\}$.
\section{Imperfection of the probe's preparation}
Even though we have shown that the optimal Fock state can give us better sensing of the coupling strength, in a real experiment, it is literally inevitable to avoid noise and imperfection during the process of state preparation. For example, imperfections in the trapped ion platform could arise from factors such as thermal noise during preparation, inaccuracies in the duration of laser pulses used in red and blue sideband interactions during Fock state preparation, and imperfect cooling of the trapped ions. In this section, we examine and investigate the fragility of the probe's sensitivity due to the experimental defects of Fock state preparation. We assume that instead of obtaining an ideal pure Fock state $\ket{n}$, we indeed prepare such Fock state with a small, but non-zero noise, which is expressed in the form of a mixed state as
\begin{align}\label{eq:a5.1}
\rho=(1-2\varepsilon)\ket{n}\bra{n}+\varepsilon\left(\ket{n+1}\bra{n+1}+\ket{n-1}\bra{n-1}\right),
\end{align}
where $\varepsilon$ represents the population of the two neighboring states $\ket{n+1}$ and $\ket{n-1}$ and $\varepsilon \ll 1$. Here, we consider a scenario where noise is equally present only in these two neighboring states, chosen for simplicity and qualitative description purposes.
The realistic state of the probe for interaction I is thus represented by 
\begin{align}\label{eq:a5.2}
\rho_{a,b,c}=&\rho_{a}\otimes \rho_{b} \otimes \rho_{c},
\end{align}
while the probe for interaction II would be
\begin{align}\label{eq:a5.3}
\rho_{a',b'}=&\rho_{a'}\otimes \rho_{b'},
\end{align}
with their subsystems in the state given in Eq.\ref{eq:a5.1}, 
\begin{align}\label{eq:a5.4}
    \rho_i = (1-2\varepsilon_i)\ket{n_i}\bra{n_i}+\varepsilon_i\left(\ket{n_i+1}\bra{n_i+1}+\ket{n_i-1}\bra{n_i-1}\right).
\end{align}
Here, the subscript $i$ denotes the motional modes associated with the trilinear interactions: $i \in {a,b,c}$ for interaction I and $i \in {a',b'}$ for interaction II. For simplicity, let us consider the case in which the noise of the interacting modes is of the same value, $\varepsilon_i=\varepsilon\ll 1$. From the tensor products of Eqs. \ref{eq:a5.3} and \ref{eq:a5.4} and the considered conditions, the probability of successfully obtaining the desired Fock states is straightforwardly the product of the probabilities of success in each mode, given as
\begin{align}
p=(1-2\varepsilon)^3.
\end{align}

\begin{figure}
    \centering
    \includegraphics[width=\textwidth]{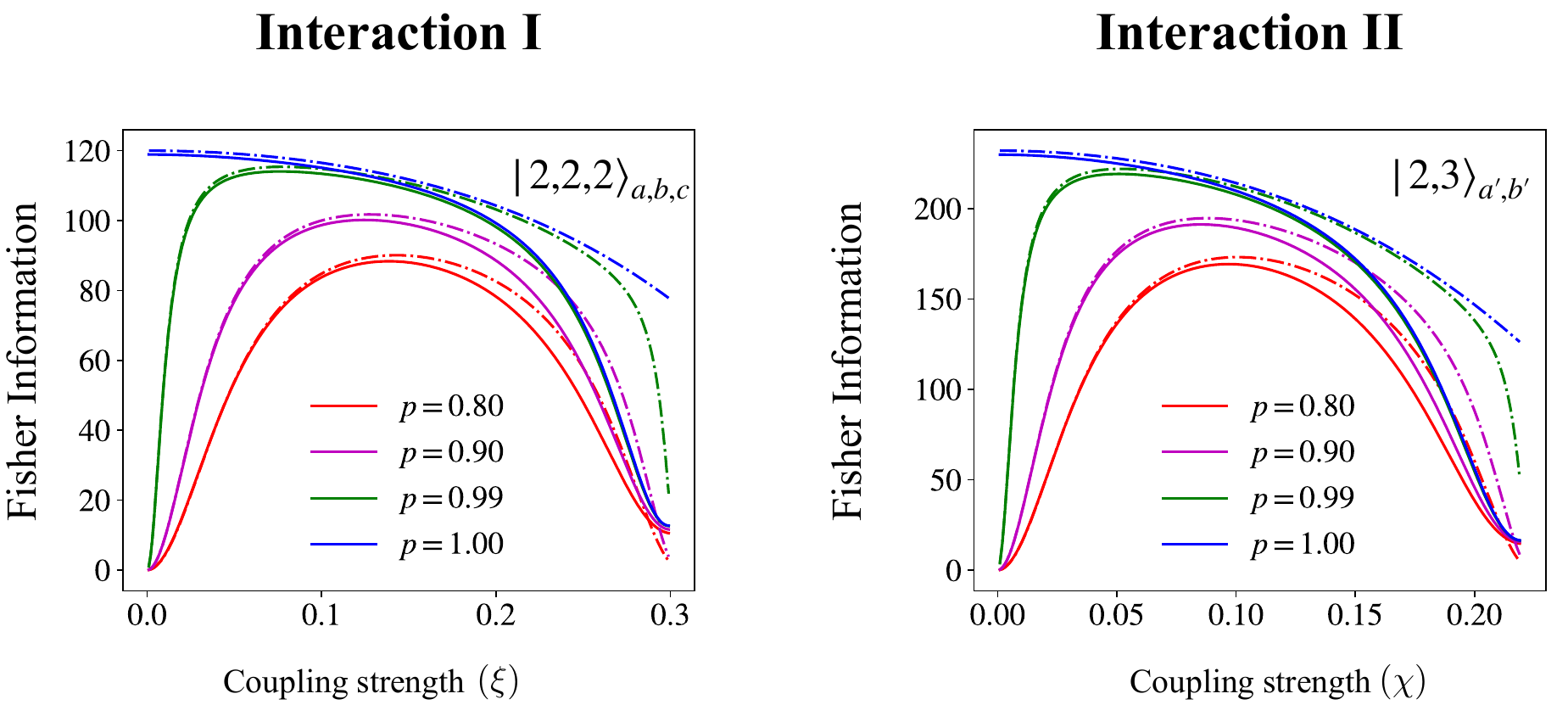}
    \caption{The comparison of Fisher information for sensing the coupling strengths $\xi$ and $\chi$ is depicted, with the probe in pure Fock states (blue) and mixed states with probabilities of successful preparation ($p = 0.80$ in red, $0.90$ in pink, $0.99$ in green). Solid lines represent Fisher information using two-shot sequential JC measurements, while dashed lines represent Fock basis measurements. For mixed states, Fisher information for sensing very small coupling strengths is near zero but increases abruptly with increasing coupling strength.}
    \label{fig:a1}
\end{figure}

Figure \ref{fig:a1} shows the Fisher information of the protocol in the case when the state of the probe is prepared in Fock states with different amounts of noise. As depicted in the figure, noise diminishes the sensitivity of the protocol for sensing near-zero coupling strengths, but it increases abruptly as the coupling strengths rise. The rate of this increase depends directly on the probability of successful preparation, denoted by $p$. The noise creates narrow troughs at the zero point in the Fisher information graphs. Moreover, the graph suggests that despite the noise introduced during the state preparation process, the probe remains sufficiently sensitive for sensing the coupling strength. In other words, the sensitivity is not completely ruined by the noise.

The reason that the noise can greatly suppress the sensitivity of the very small coupling strengths $\xi\rightarrow 0$ and  $\chi\rightarrow 0$ can be analyzed from the form of the classical Fisher information. From Eq.\ref{eq:a1.1}, each term in the summation of the Fisher information is obtained by the ratio of the changing rate of the probabilities due to the sensing parameter and the probabilities themselves. In the case of pure Fock states $\ket{n}$ the populations of the neighboring states $\ket{n\pm 1}$ are zero by definition. A slight deviation from zero population in these neighboring states would be easily notice, giving a substantial amount of Fisher information indicative of the nonlinear interaction. In contrast, mixed states initially have populations in the neighboring states $\ket{n\pm 1}$, resulting in less distinct changes in phonon statistics due to nonlinear interactions compared to Fock states.

\section{Sensitivity of Coherent States}
In order to perceive the optimal sensitivity obtained from the probe in Fock states, we have to compare it with the performance provided by a classical state. Here we choose coherent states as a representative of classical states and treat it as a benchmark. They are one of the most well-known classical states in quantum optics describing the light produced by a stabilized laser.  
In the case that the probe is prepared in a coherent state $\ket{\alpha,\beta,\gamma}_{a,b,c}$ for sensing interaction I and $\ket{\alpha',\beta'}_{a',c'}$ for interaction II, as discussed in section \ref{sec:ap.2}, the QFI is determined from the variances of the generators $\hat{G}_{\rm I}$ and $\hat{G}_{\rm II}$.
The QFI of a coherent states $\ket{\alpha,\beta,\gamma}_{a,b,c}$, represented by $\mathcal{F}^{\rm co}_{\rm Q,I}$, for interaction I is given as
\begin{align}
\mathcal{F}^{\rm co}_{\rm Q,I}&=4t^2\left(|\alpha|^2|\beta|^2+|\alpha|^2|\gamma|^2+|\beta|^2|\gamma|^2+|\alpha|^2\right)\nonumber \\
&= 4t^2(n_a n_b+n_a n_c+ n_b n_c+n_a),
\end{align}
where $n_a=|\alpha|^2$, $n_b=|\beta|^2$ and $n_c=|\gamma|^2$ are the mean phonon numbers of modes $a$, $b$ and $c$ respectively.
For interaction II, the QFI of a coherent state $\ket{\alpha',\beta'}_{a',c'}$ is represented by
\begin{align}
\mathcal{F}^{\rm co}_{\rm Q,II}&=4t^2\left(|\beta'|^4+3|\alpha'|^2|\beta'|^2+2|\alpha|^2\right)\nonumber \\
&= 4t^2(n^2_{b'}+3n_{a'}n_{b'}+2n_{a'}).
\end{align}

The QFIs provides an upper bound on Fisher information, but in practical applications, the achieved sensitivity of coherent states is often much lower than their QFIs. For a fair comparison, we employ an identical sensing protocol as depicted in Figure \ref{fig:2} of the main text, where measurements are performed on the qubit's state.

\begin{figure}
    \centering
    \includegraphics[width=\textwidth]{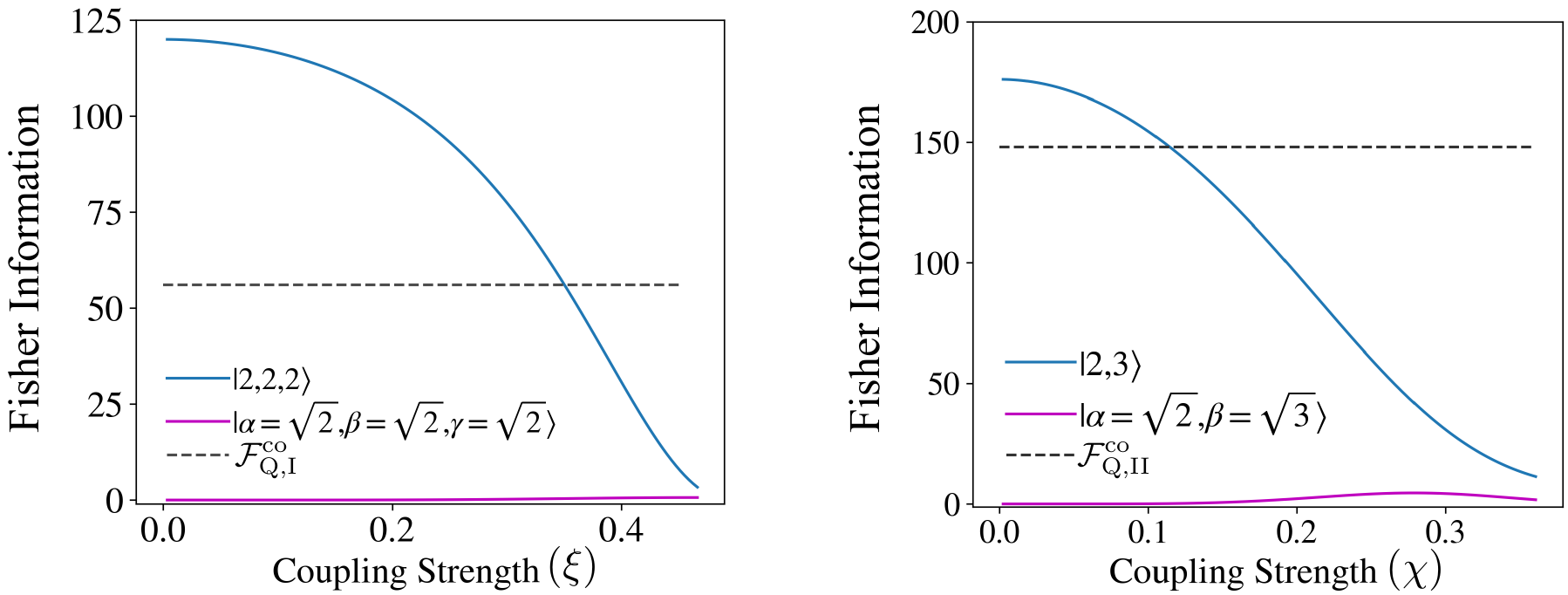}
    \caption{({\bf a}) The figure demonstrates the Fisher information for sensing the coupling strength $\xi$ of interaction I when the probe is initially set in an optimal Fock state $\ket{2,2,2}$ (blue), compared to the case of coherent states (pink) with the same excitation number in each motional mode. ({\bf b}) The Fisher information for sensing the coupling strength $\chi$ of interaction II with the probe in $\ket{2,3}$ (blue) is compared to the case of coherent states (pink) with the same excitation number. The dashed lines represent the quantum Fisher information of the coherent states.}
    \label{fig:a2}
\end{figure}

In Figure \ref{fig:a2}, we compare the sensitivity of the protocol using Fock states and coherent states. It's evident that the sensitivity of the coherent state is significantly lower than that of Fock states, even falling short of their QFIs. This discrepancy arises because changes in the state are less inferable from qubit measurements compared to Fock states, where slight changes in population lead to different probabilities of qubit excitation \cite{Ritboon2022}. This limitation also applies to thermal states with the same average excitation. Moreover, achieving accurate estimations of coupling strength requires precise controls and maneuvers for both coherent and thermal state preparations \cite{Leibfried2003}.

\bibliographystyle{plain}

\end{document}